\documentclass[longauth,letter]{aa} %
\usepackage{graphicx}
\usepackage{txfonts}
\usepackage{csvsimple,booktabs}

\usepackage{tablefootnote}

\usepackage{natbib}
\usepackage[colorlinks=true,allcolors=blue]{hyperref} 
\usepackage{multirow}
\usepackage{lastpage} %to fix the page count

\bibpunct{(}{)}{;}{a}{}{,}             

\newcommand*{\teff}{\ensuremath{T_{\text{eff}}}}
\defcitealias{squicciarini25}{S25}

\begin{document}

   \title{GPI+SPHERE detection of a $6.1 M_{\text{Jup}}$ circumbinary planet around HD 143811}
 %  \subtitle{subtitle}

   \author{V. Squicciarini\inst{1,2}             \and J. Mazoyer\inst{2}
            \and C. Wilkinson\inst{2}
            \and A.-M. Lagrange\inst{2,3}
            \and P. Delorme\inst{3}
            \and A. Radcliffe\inst{2}
            \and O. Flasseur\inst{4}
            \and F. Kiefer\inst{2}
            \and E. Alecian\inst{3}
          }

   \institute{
        Department of Physics \& Astronomy, University of Exeter, Stocker Road, Exeter, EX4 4QL, UK; \\
        \email{v.squicciarini@exeter.ac.uk}
        \and
        LIRA, Observatoire de Paris, Université PSL, Sorbonne Université, Université Paris Cité, CY Cergy Paris Université, CNRS, F-92190 Meudon, France
        \and
        Univ. Grenoble Alpes, CNRS-INSU, Institut de Planétologie et d'Astrophysique de Grenoble (IPAG) UMR 5274, Grenoble, F-38041, France
        \and
        Centre de Recherche Astrophysique de Lyon (CRAL) UMR 5574, CNRS, Univ. de Lyon, Univ. Claude Bernard Lyon 1, ENS de Lyon, F-69230 Saint-Genis-Laval, France
    }

   %\date{Received September 15, 1996; accepted March 16, 1997}

    \abstract
% 5 {} token are mandatory
    {Owing to its sensitivity to wide-orbit giant exoplanets, direct imaging is uniquely positioned to shed light on the interplay between protoplanetary disks and stellar hosts. In addition to constraining formation models, new detections are natural benchmarks for an atmospheric characterization.}
    {The COBREX project performed an extensive reanalysis of archival observations from SPHERE and GPI using advanced post-processing techniques, that enhanced the detection sensitivity at close separation. Newly found companion candidates are being followed up to confirm new planets.}
    {Following the detection of a companion candidate around the young ($\sim 15$ Myr) binary star HD 143811, we collected a new observation with SPHERE@VLT (0.95-1.67 $\mu$m) to confirm the presence of the source and to assess its physical bond to the target.}
    {We report the discovery of a new exoplanet orbiting HD 143811 at a projected separation of 0.43" $\sim 60$ au. Based on a 9-year-long baseline, we derive a mostly face-on and low-eccentricity orbit with a period of $320 ^{+250}_{-90}$ years. The luminosity of the planet, constrained through the H-band spectrum from GPI, H-band photometry from SPHERE/IRDIS and YJ upper limits from SPHERE/IFS, allows us to place strong constraints on the intrinsic temperature of the planet (T$_{\text{int}} = 1000 \pm 30$K), which corresponds to a mass of $6.1^{+0.7}_{-0.9} ~M_{\text{Jup}}$.}
    {HD 143811(AB)b is the second planet ever discovered by GPI. It joins the small cohort of circumbinary planets discovered through imaging and becomes a prime target for follow-up formation, dynamical, and characterization studies.}


\titlerunning{}

\maketitle
%
%-------------------------------------------------------------------

\section{Introduction}
Understanding the physical processes of planet formation and the intricate chemical and dynamical interplay of planets with parent disks that is reflected in their internal structure and in their atmosphere is a major challenge in modern astronomy. Compared to single stars, multiple systems provide additional challenges for planet formation \citep[see][and ref. therein]{fontanive2021,thebault25}. Unlike transit and radial velocities, direct imaging (DI) probes wide-orbit giant planets around young ($t \lesssim 1$ Gyr) stars, which offers a complementary window on planet formation \citep{gaudi21}. Circumbinary planets are particularly compelling targets for DI: their disks evolve under nonstandard initial conditions and their planets may either form in situ at large separations or be dynamically scattered outward \citep{nelson03,pierens08,marzari19,gianuzzi23}. Despite dedicated searches \citep[e.g.][]{asensiotorres18}, only a few circumbinary planets have been imaged to date (HD 106906 b, \citealp{Bailey2014_hd106906}; b Cen b, \citealp{janson21_bcen}; Ross 458 b, \citealp{goldman2010}; DLR 1 b, \citealp{delorme2013}).

The Gemini Planet Imager (GPI) was a second-generation high-contrast imaging instrument on the Gemini South Telescope that was active from 2014 to 2020 \citep{macintosh2014_SPIE_firstlight}. The GPI Exoplanet Survey (GPIES) was a 5-year blind survey of 600 nearby young stars. It was the largest such survey conducted to date and resulted in the discovery of one new exoplanet \citep[51 Eri b,][]{macintosh15} and one new brown dwarf \citep[HR 2562 B,][]{konopacky16}. Together with the 400-stars SpHere INfrared survey for Exoplanets (SHINE) conducted with the Spectro-Polarimetric High-Contrast Exoplanet Research \citep[SPHERE,][]{beuzit19} at the Very Large Telescope, it provided the tightest statistical constraints to the occurrence of wide-orbit exoplanets \citep{nielsen19, vigan21}. In the past few years, new post-processing algorithms have been developed to enhance the signal-to-noise ratio (S/N) of existing observations, which opened up a new detection window in the parameter space in which previously undetected planets might reside. This is the main goal of the project caleld COupling data and techniques for BReakthroughs in EXoplanetary systems exploration (COBREX), which undertook a complete re-reduction of SHINE \citep{chomez25} and GPIES \citep[][hereafter S25]{squicciarini25} using an algorithm named PAtch-COvariance \citep[PACO,][]{Flasseur_paco}.

The GPIES survey reanalysis by \citetalias{squicciarini25} achieved deeper detection limits than were previously obtained on a 150-star subsample \citep{nielsen19}, with up to a two-fold gain in minimum detectable mass. This allowed us to identify new promising candidates. One of these, detected around the star HD 143811 = HIP 78663, was observed twice (in 2016 and 2019) and showed indications of a common proper motion. The $\sim 3.5 \sigma$ detection in the 2019 observation did not meet our strict $5 \sigma$ detection threshold that is required for confirmation, however. For this reason, we proposed it to be re-observed with SPHERE. In this letter, we report the confirmation of this candidate as a planet, the first GPI + SPHERE joint exoplanet detection.

\section{The system}\label{sec:star}

HD 143811 belongs to the young (5-30 Myr) Scorpius-Centaurus, the nearest OB association \citep{de_zeeuw99}. Its membership is confirmed by BANYAN $\Sigma$ \citep{gagne18} using Gaia DR3 \citep{gaia_dr3} data. 

The star is classified as a spectroscopic binary in the literature \citep{zakhozhay_RVbinary_2022, grandjean_harpsbinary_2023}. Although too close ($< 1$ au $\sim 7$ mas) to be detected in DI, the astrometric
solution from Gaia shows the presence of a companion at a 2.8 $\sigma$ level (Appendix~\ref{app:star_params}). The presence of an unresolved companion demands a dedicated fitting of the measured photometry to determine the age of the system and the two masses. The main stellar parameters are listed in Table~\ref{tab:star_data}; the masses and effective temperatures are to be read as ranges encompassing the random and systematic (i.e., model-dependent) uncertainties. We refer to Appendix~\ref{app:star_params}, where we describe the derivation of these parameters and their uncertainties.

{\centering
\begin{table}[t!]
\caption{Main astrometric, kinematic, photometric and astrophysical properties of the system.}\label{tab:star_data}
 {\small 
\begin{tabular}[l]{l|l|l|l|l|l}
\hline\hline
 & \textbf{Value} & \textbf{Ref.} & & \textbf{Value} & \textbf{Ref.} \\\hline
$\alpha$ & 240.8892$^\circ$ & 1 & H & $7.784(47)$ mag & 2\\
$\delta$ & -30.1372$^\circ$  & 1 & age & $18 \pm 3$ Myr & 3 \\
$d$ & $136.86(38)$ pc & 1 & $M_1$ & $1.24 - 1.40~M_\odot$ & 3 \\
$\mu_{\alpha}^*$ & $-14.850(26)$ mas/yr & 1 & $M_2$ & $1.08 - 1.28 ~M_\odot$ & 3 \\
$\mu_{\delta}$ & $-24.984(16)$ mas/yr & 1 & $T_{\text{eff},1}$ & $6250 - 6900$ K & 3\\
G & $8.798(3)$ mag & 1 & $T_{\text{eff},2}$ & $5750 - 6500 $ K & 3
\\
 \hline
\end{tabular}
}
\tablefoot{$\alpha$, $\delta$: RA, dec (J2016). $d$: distance. $\mu_{\alpha}^*$, $\mu_{\delta}$: proper motion. $M_1$, $M_2$, $T_{\text{eff},1}$, $T_{\text{eff},2}$: stellar masses and eff. temperatures. References: 1: \citet{gaia_dr3}; 2: \citet{2mass}; 3: this work.}
\end{table}
}

\section{Observations and data reduction}\label{sec:data}

Observed in the GPI H-band (1.50-1.78 $\mu$m) as part of GPIES in 2016 and 2019, HD 143811 was classified as an interesting target for follow-up by \citetalias{squicciarini25} after the detection of a dim source (S/N$=5.4$) at about 0.4 arcsec away from the central binary. The S/N of the second epoch was low (S/N$=3.5$), and we were therefore unable to confirm the candidate (hereafter, CC1) as a real companion. For these reasons, we requested a third epoch with SPHERE, which was obtained on July 24, 2025 (Program ID 114.27LG.002). The SPHERE observation was carried out in the IRDIFS mode: the Integral Field Spectrograph \citep[IFS;][]{claudi08} collected YJ-band (0.95-1.33 $\mu$m) spectroscopy over a $1.7" \times 1.7"$ field of view (FoV), whereas the Infra-Red Dual-band Imager and Spectrograph \citep[IRDIS;][]{dohlen08} provided dual-band imaging in H23 ($1.593~\mu$m, $1.667~\mu$m). The main details of the observations are provided in Table~\ref{tab:obs}. A complete description of the pipeline that handles the data reduction from raw frames to PACO S/N map can be found in \cite{chomez23} for SPHERE data and in \citetalias{squicciarini25} for GPI data.

\begin{table}[t!]
\caption{Details of the observations of HD 143811.}\label{tab:obs}
\vspace{-7mm}
\begin{center}
 {\small 
\begin{tabular}{cccccc}
\hline
\hline
Date (UT) & Instrument & Int. time  & Seeing& $\tau_0$ &  $\Delta$PA ($^{\circ}$) \\
 & & (min) & (arcsec) & (ms) & \\
 
\hline 

2016/04/30  & GPI & 43.7 & 2.24
 & N.A. & 174.4 \\
2019/08/10  & GPI & 47.7 & N.A. & N.A. & 174.8 \\
2025/07/24 & SPHERE & 90 & 0.33 & 4.8 & 52.1 \\   
\hline
\end{tabular}
}
\end{center}
\vspace{-5mm}
\tablefoot{$\Delta$PA: total FoV rotation. Coherence time and seeing are averaged over the exposures. N.A.: values that could not be found.}
\end{table}

\section{Results}\label{sec:results}

\subsection{Confirmation of the planet}

\begin{figure*}[t!]
    \centering
    \includegraphics[trim={0cm 1cm 0cm 0.5cm},clip,width=\linewidth]{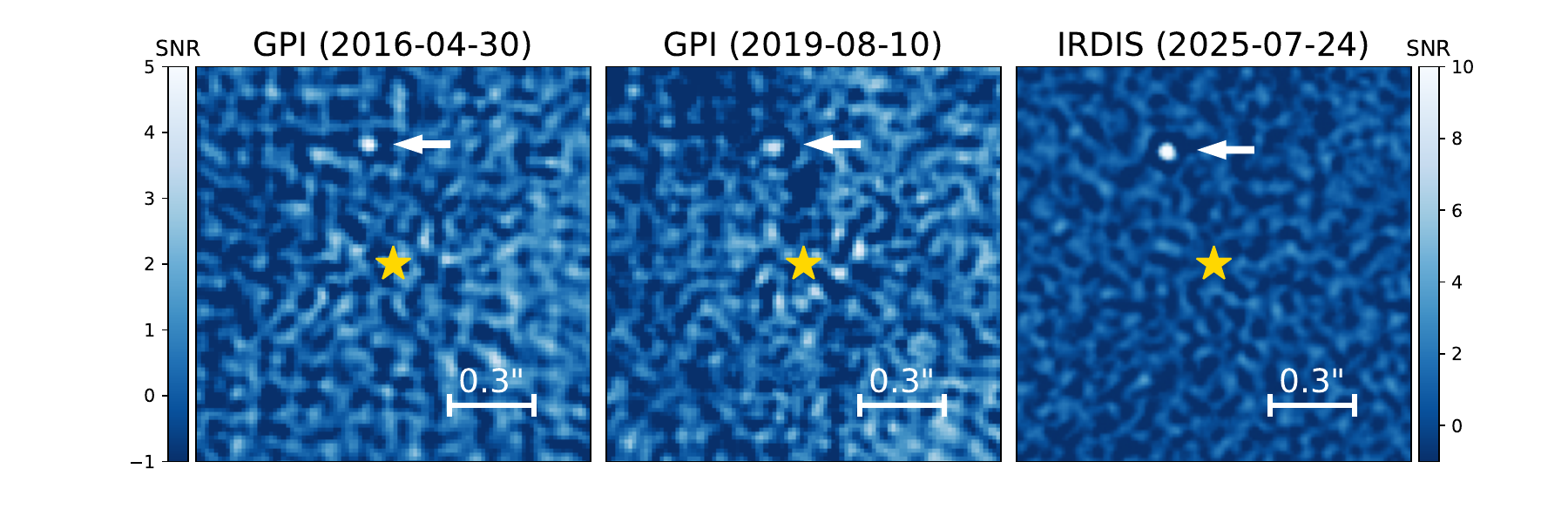}
    \caption{S/N maps produced from PACO in the three epochs. The image scale is shown in the lower right corner, and the position of the planet is indicated by an arrow. The left color bar refers to the two GPI maps.}
    \label{fig:paco_images}
\end{figure*}

The S/N maps from PACO for the three epochs are reported in Figure~\ref{fig:paco_images}. CC1 is clearly redetected (S/N$=17.8$) at the expected position in IRDIS data (sep $=431.1\pm0.63$ mas, PA $=21.71\pm0.09$ deg); conversely, no detection could be found in IFS data. The details of the astrometry and the photometry of the source are provided in Table~\ref{tab:detections}. The good agreement between GPI (2016) and SPHERE/IRDIS ($H2$, $H3$) photometry\footnote{SPHERE H2- and H3-band photometry was estimated from GPI data as described in \citetalias{squicciarini25}.} is clear; we consider the GPI (2019) photometry as unreliable due to the low S/N.

\begin{table*}[t!]
\begin{center}
\caption{Astrometric and photometric properties of CC1. H2 and H3 indicate the absolute magnitude in the IRDIS H23 bands.}\label{tab:detections}
 {\small 
\begin{tabular}{cccccccc}
\hline
\hline
ObsID & S/N & Contrast & Separation  & PA & $H2$ & $H3$ & $H2$-$H3$ \\
 & & (mag) & (mas) & ($^{\circ}$) & (mag) & (mag) & (mag) \\ 
\hline 

GPI H (2016) & 5.4 & 11.9 & $427.8 \pm 3.0$ & $11.79\pm0.46$ & $14.14 \pm 0.32$ & $13.85 \pm  0.29$ & $0.29 \pm  0.40$ \\
GPI H (2019) & 3.5 & 12.5 & $427.3 \pm 3.3$ & $16.29 \pm 0.38$ & $15.00 \pm  0.25$ & $14.27 \pm 0.19$ &  $0.73 \pm  0.27$ \\
SPHERE H23 (2025) & 17.8 & 12.0 & $431.1 \pm 0.6$ & $21.71\pm0.09$ & $14.31 \pm 0.10$ & $13.85 \pm 0.10$ & $0.44 \pm 0.14$ \\
\hline

\end{tabular}
}
\end{center}
\end{table*}

As is usual in DI, the confirmation of CC1 involved the comparison of its astrometric displacements across the three epochs with the motion expected from a background interloper. As a first step, we verified that the relative proper motion of the source was compatible with a bound orbit around the central binary. We then constructed a suitable sample of synthetic background stars using the Besançon Galaxy Model \citep{czekaj2014} and found that a proper motion such as the one of the source is ruled out at a C.L. = 3.6 $\sigma$.
An additional argument against the background scenario is provided by the nondetection of the source in IFS data, which sets a lower limit to the $J-H2$ color of the source, $J-H2 > 1.4$ mag. While a red color like this is not atypical for young giant planets \citep[see, e.g., HR 8799 planets;][]{bonnefoy16}, we found that no simulated source appears to be as red. In other words, every source from the synthetic sample would have been detected by IFS.

Based on these two lines of evidence (described in more detail in Appendix~\ref{app:proper_motion}), we rule out the hypothesis of a background contaminant. We confirm that CC1 (hereafter, HD 143811 (AB)b) is part of the HD 143811 system.

\subsection{Orbital properties}
In order to constrain the orbital parameters of the planet, we used the {\tt orbitize!} v3 package \citep{blunt_orbitize_2024}, specifically, the Orbits for the Impatient (OFTI) algorithm \citep{blunt_orbits_2017}. We included the three astrometric points (GPI IFS 2016, GPI IFS 2019, and SPHERE IRDIS 2025) in Table \ref{tab:detections} and used a binary star mass of $M=2.44 \pm 0.09~M_\odot$ (see Appendix~\ref{app:star_params}) and the parallax from Table~\ref{tab:star_data}. {\tt orbitize!} default priors were used. The OFTI algorithm was run until we obtained $10^6$ accepted orbits. The planet astrometry is consistent with a long-period orbit ($T \sim 300$ yr, corresponding to $a \sim 60$ au) with low eccentricity ($e=0.19 \pm 0.12$) that is seen almost face-on ($i=37^{\circ+11}_{-12}$). The impact of systematic stellar mass uncertainties is negligible ($\Delta a/a<2\%$, $\Delta e/e<10\%$, $\Delta i/i<5\%$, and $\Delta T/T<0.3\%$). A randomly drawn sample of orbits extracted from the posterior are plotted in their sky projection in Fig~\ref{fig:orbitize}. Additional details on the priors and the posteriors can be found in Appendix~\ref{app:orbitize}.

\begin{figure}[t!]
    \centering
    \includegraphics[width=\linewidth]{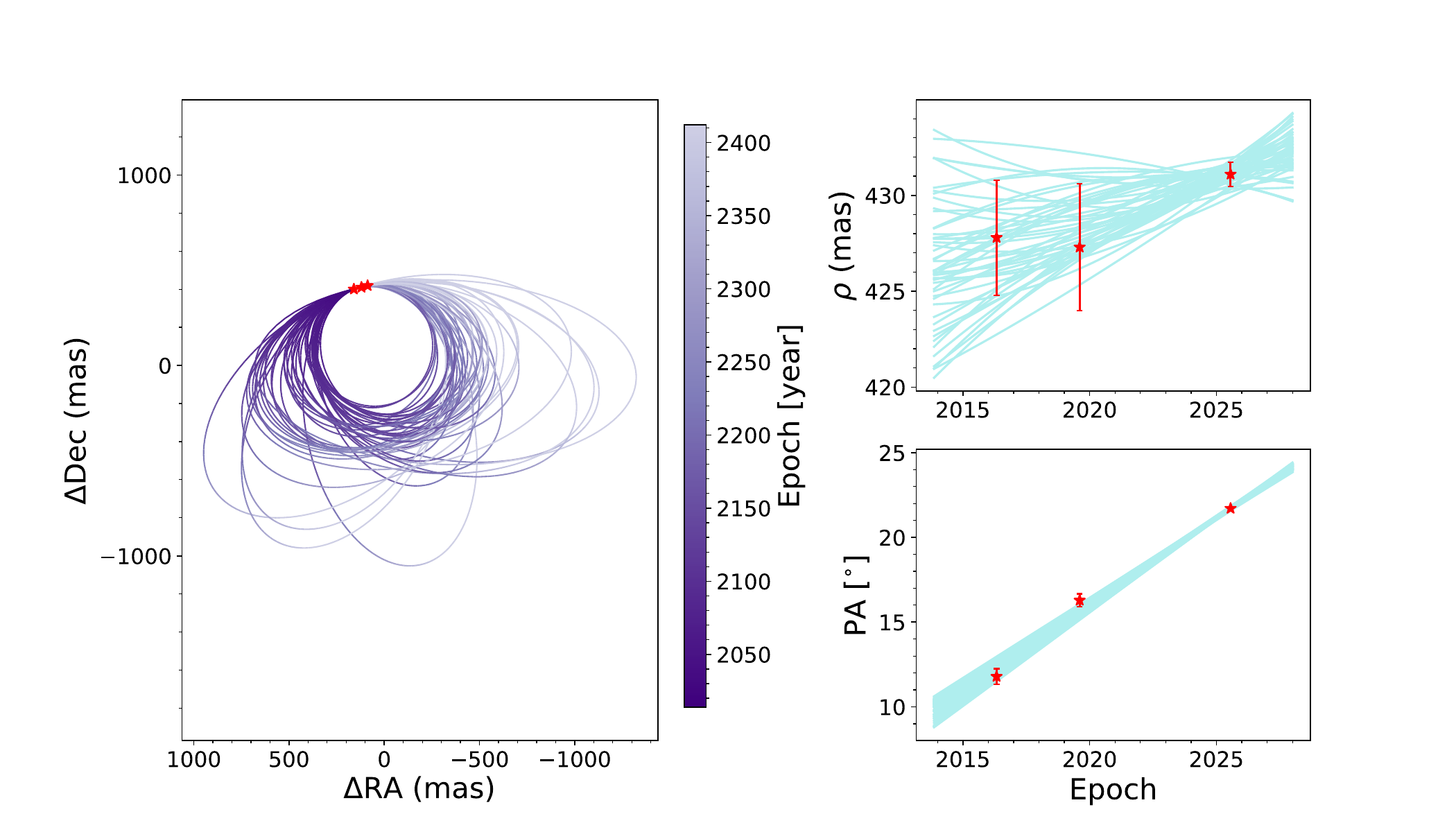}
    \caption{Orbits of HD 143811 b. Left: 50 randomly drawn accepted orbits from the OFTI posterior plotted in their sky projection. The measured astrometry is plotted in red from GPI (2016 and 2019) and SPHERE/IRDIS (2025) data. Right: same randomly drawn accepted orbits plotted as a function of time. The red dots show the measured astrometry and associated errors.}
    \label{fig:orbitize}
\end{figure}

\subsection{Physical properties}\label{sec:atmo_fit}

The photometric data from Table~\ref{tab:detections} can be used to constrain the physical properties of the planet by comparison with atmospheric models. In the case of IFS, the wavelength-collapsed detection limit from PACO (enhancing the sensitivity by a factor $\sim \sqrt{n_{wvl,IFS}} = \sqrt{39}$ compared to individual $\lambda$ slices) at the expected source separation was used as upper limits, adopting an effective $\lambda_{IFS} = 1.14 \pm 0.10 ~ \mu$m. The data from 2019 were not used because the S/N is too low. The procedure to convert contrast data into absolute flux measurements was described in \citetalias{squicciarini25}, and the slight variations are detailed in Appendix~\ref{app:physical_fit}. A binary template was adopted for the star; the impact of the mass ratio and $T_{\text{eff}}$ uncertainties on the derived planet flux is negligible.

The spectrophotometric data were analyzed using a grid of atmosphere-interior models from the HADES model \citep{wilkinson24}. Six physical parameters were adjusted within a Monte Carlo Markov chain (MCMC) framework. We adopted uniform (flat) priors for all fitted parameters as listed in Table \ref{tab:inferred_params_atmo}. The age of the system was used as an additional constraint and was predicted from the other parameters by adopting a medium entropy value per mass from \cite{mordasini17}, which is equivalent to a warm start. The upper limit for the system age ($21$ Myr) was treated as an upper limit, while a lower limit was set based on the estimated median disk life for an F-type star \citep[$\tau = 5$ Myr;][]{pfalzner22}. Details of the implementation of the MCMC, of the age limits, and of the derived posteriors can be found in Appendix~\ref{app:physical_fit}.

\begin{figure}[t!]
    \centering
    \includegraphics[width=\linewidth]{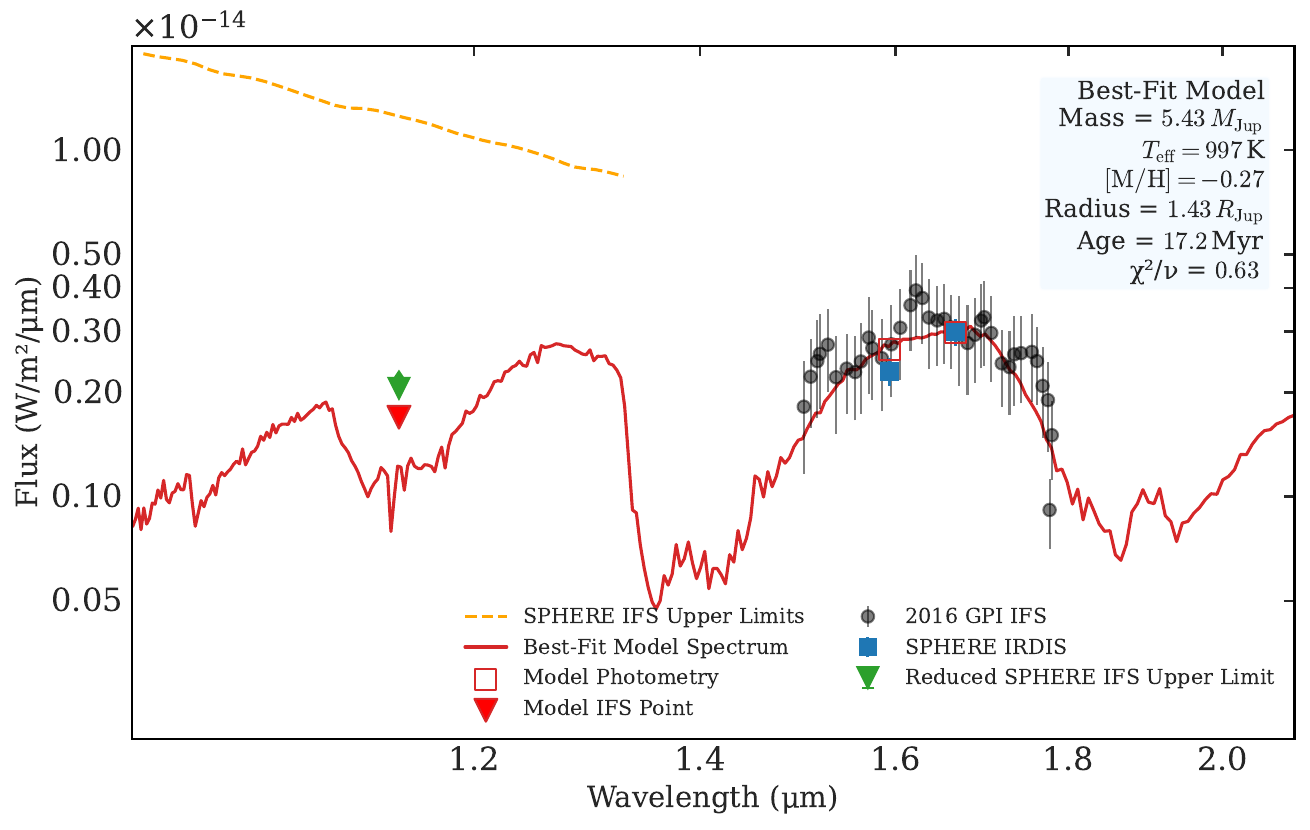}
    \caption{Spectrophotometry of the planet, normalized to $d=10$ pc, compared to the best-fit model from HADES (red line). The black circles show the GPI spectrum, the blue square shows SPHERE photometry, and the green arrow indicates the SPHERE-IFS $5 \sigma$ upper limit (all with 1$\sigma$ errors). The orange line shows $5 \sigma$ IFS upper limits in individual $\lambda$ slices. The predicted photometric points are shown with red markers.}
    \label{fig:best_fit_spectrum}
\end{figure}

The best-fitting sample, with the highest log-likelihood, is shown in Figure \ref{fig:best_fit_spectrum}. The GPI and SPHERE photometry are fit well by the model. This result also explains, without any tension within the fit, why there is a nondetection with SPHERE-IFS.
Our analysis provides a strong constraint on the intrinsic temperature of the planet: $T_{\mathrm{int}} =1000 \pm 30\,\mathrm{K}$. Combined with the constraint on the system age, this temperature implies a mass of $6.1^{+0.7}_{-0.9}\,M_{\mathrm{Jup}}$. The radius, derived from the other physical parameters at each MCMC sample, is $1.4\pm0.1\,R_{\mathrm{Jup}}$. The atmospheric metallicity is weakly constrained around a solar value, while the core mass and sedimentation efficiency ($f_{\mathrm{sed}}$) remain poorly constrained. 
The fit also indicates vigorous vertical mixing, although this parameter is weakly constrained. If this finding is confirmed, it would suggest an atmosphere in chemical disequilibrium.

\section{Discussion}\label{sec:discussion}

By confirming the candidate that was first identified by \citetalias{squicciarini25}, the robust (S/N=18) detection in the new IRDIS data allowed us to unambiguously establish its status as a planetary companion to the HD 143811 system. The nondetection in IFS data is consistent with the flux level expected for a $\teff=1000$ K object according to combined atmosphere-interior models at the age and distance of the host system; moreover, the GPI spectrum and the IRDIS photometry show excellent agreement, which highlights the robustness of the estimated fluxes. The proper motion of the source is clearly distinguished from the proper motion expected from background sources, and is is indicative of orbital motion around the central binary.

This discovery adds an important data point for comparative exoplanetology, bridging the temperature range between AF Lep b \citep{derosa23, mesa23, franson23} and cooler planets with a slightly lower mass, such as 51 Eri b \citep{macintosh15}. Together, these directly imaged planets of varying ages and masses offer a unique testbed for models of giant planet cooling and evolution. Moreover, HD 143811 b joins the small sample of circumbinary planets detected in imaging. Monitoring the binary with radial velocities and interferometry, along with astrometric orbit follow-up, will constrain the dynamical history of  the system. While a characterization of the atmosphere is currently limited by the spectral coverage, broader wavelength and high-resolution observations especially in the mid-infrared will be essential for probing temperature structure and composition. This characterization will clarify the differences between circumbinary and single-star planets and their relation to free-floating objects \citep[e.g.][]{sutherland16,coleman24}.

\begin{acknowledgements}
We are grateful to C. Babusiaux for the discussion about the reliability of isochronal effective temperatures.
This project has received funding from the European Research Council (ERC) under the European Union's Horizon 2020 research and innovation program (COBREX; grant agreement n° 885593).
SPHERE is an instrument designed and built by a consortium consisting of IPAG (Grenoble, France), MPIA (Heidelberg, Germany), LAM (Marseille, France), LESIA (Paris, France), Laboratoire Lagrange (Nice, France), INAF - Osservatorio di Padova (Italy), Observatoire de Genève (Switzerland), ETH Zürich (Switzerland), NOVA (Netherlands), ONERA (France) and ASTRON (Netherlands) in collaboration with ESO. SPHERE was funded by ESO, with additional contributions from CNRS (France), MPIA (Germany), INAF (Italy), FINES (Switzerland) and NOVA (Netherlands). SPHERE also received funding from the European Commission Sixth and Seventh Framework Programmes as part of the Optical Infrared Coordination Network for Astronomy (OPTICON) under grant number RII3-Ct-2004-001566 for FP6 (2004-2008), grant number 226604 for FP7 (2009-2012) and grant number 312430 for FP7 (2013-2016). This work has made use of the High Contrast Data Centre, jointly operated by OSUG/IPAG (Grenoble), PYTHEAS/LAM/CeSAM (Marseille), OCA/Lagrange (Nice), Observatoire de Paris/LESIA (Paris), and Observatoire de Lyon/CRAL, and is supported by a grant from Labex OSUG@2020 (Investissements d’avenir - ANR10 LABX56).
This work is based on observations obtained at the Gemini Observatory, operated by the Association of Universities for Research in Astronomy, Inc., under a cooperative agreement with the NSF on behalf of the Gemini partnership: the National Science Foundation (United States), the National Research Council (Canada), CONICYT (Chile), the Australian Research Council (Australia), Ministério Ciência, Tecnologia e Inovação (Brazil) and Ministerio de Ciencia, Tecnología e Innovación Productiva (Argentina).
This research has made use of SIMBAD,  VizieR, CDS, Strasbourg Astronomical Observatory, France.
This work is supported by the French National Research Agency in the framework of the Investissements d’Avenir program (ANR-15-IDEX-02), through the funding of the “Origin of Life” project of the Univ. Grenoble-Alpes.
This work was granted access to the HPC resources of MesoPSL financed by the Region Ile de France and the project Equip@Meso (reference ANR-10-EQPX-29-01) of the programme Investissements d’Avenir supervised by the Agence Nationale pour la Recherche.
\end{acknowledgements}

\bibliographystyle{aa}
\bibliography{bibliography.bib}

\begin{appendix}

\section{Stellar parameters}
\label{app:star_params}

According to BANYAN $\Sigma$, HD 143811 has a 70\% membership probability to Upper Scorpius (US) and a 30\% to Upper Centaurus-Lupus (UCL), two of the three regions in which Scorpius-Centaurus is classically divided \citep{de_zeeuw99}. Indeed, the sky coordinates of the system are intermediate between the two subgroups, placing it in a region whose typical age is 13-15 Myr \citep{pecaut16}.

As in \citetalias{squicciarini25}, we estimated the reddening $E(B-V)$ by integrating the map by \citet{leike20} along the line of sight, finding $E(B-V) = 0.044$ mag. A larger value, $E(B-V) = 0.11$ mag, is suggested by \citet{gontcharov2017}; we allowed thus the parameter to vary between these two extremes.

The derivation of constraints on the mass ratio between the two stars is complicated by the lack of a direct detection of the secondary. As a first step, we used GaiaPMEX \citep{kiefer24a}, a tool featuring a model of the Renormalized Unit Weight Error (\texttt{ruwe}) and of the Gaia-Hipparcos proper motion anomaly \citep{kervella22}) expected for single stars, to try to constrain the properties of the companion. The \texttt{ruwe}, significant at a 2.8$\sigma$ level, and the nondetection via PMa constrain the companion to a semi-major axis shorter than 10 au, and more probably < 3 au. However, no constraint could be found on the mass ratio owing to the large degeneracy between semi-major axis and companion mass. Fortunately, the SB2 signature detected by \citet{grandjean_harpsbinary_2023} provides strong evidence for a flux ratio $\gtrsim 0.5$, corresponding to a mass ratio $q \gtrsim 0.9$.

We decided to undertake a more precise derivation of the mass ratio of the two components by exploiting the 5 spectra acquired by HARPS \citep{harps}. The spectra are separated in two main epochs (JD-2458243 and JD-2458606). Following the methodology explained in \citet{kiefer18}, the spectra were fed to TODCOR \citep{mazeh94}, a tool that calculates a 2D cross correlation function based on a linear combination of two synthetic spectra and an observed spectrum.

Synthetic spectra were taken from the PHOENIX library \citep{husser13}. The stellar parameters defining the two synthetic spectra from the library used as input for TODCOR were optimized through an iterative $\chi^2$ minimization, iteratively fitting the stellar parameters ($T_{\text{eff, j}}$, $\log{g_j}$, $(V\sin(i))_j$, where j=A,B; $\alpha=F_B/F_A$ the flux ratio at 5000 Å; [Fe/H] was set to solar) one after the other. We used a single order at $\sim 610$ nm (order 61 in HARPS) that contains many different, well separated, deep as well as shallow, absorption lines. The two SB2 components are well separated in all the 5 spectra of HD 143811. The five HARPS spectra were fitted independently, taking in the end the average and standard deviation of each parameter as our best-fit value.

\begin{figure}[h!]
    \centering
    \includegraphics[width=1\linewidth]{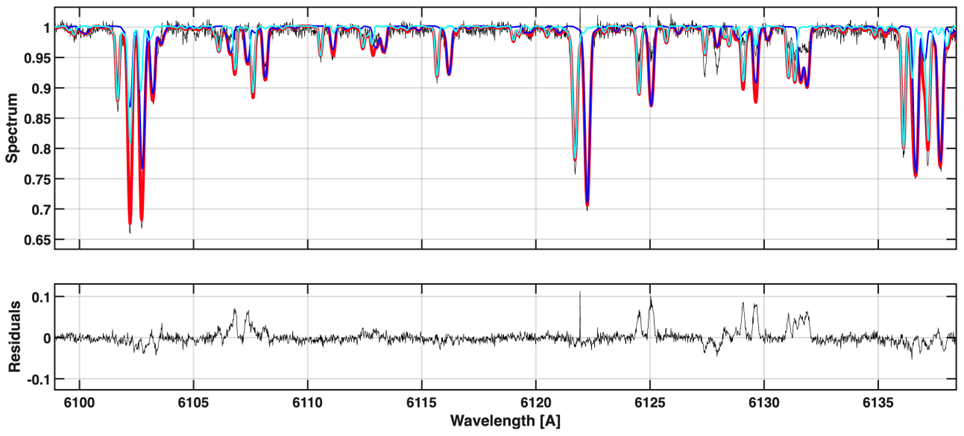}
    \caption{Model (red line) vs. observed (black line) spectrum compared in order 61. The dark blue and the cyan lines indicate the models of two individual components. If several of the lines, especially the deepest ones are well fitted, some other do not correspond well to the model.
    }
    \label{fig:todcor_1}
\end{figure}

The fit, shown in Figure~\ref{fig:todcor_1}, yields $T_{\mathrm{eff},A} = 6791 \pm 100 \,\mathrm{K},~ 
\log g_A = 4.34 \pm 0.01,~ 
v_A \sin i = 7.5 \pm 0.5 \,\mathrm{km\,s^{-1}}$, $T_{\mathrm{eff},B} = 6253 \pm 150 \,\mathrm{K}, ~
\log g_B = 4.41 \pm 0.02, ~
v_B \sin i = 5.0 \pm 0.6 \,\mathrm{km\,s^{-1}}, ~ \alpha = 0.77 \pm 0.05$. If we assume that $F_B/F_A \sim q^4$ in the visible\footnote{Lindegren et al. 2022 technical note: \url{https://dms.cosmos.esa.int/COSMOS/doc_fetch.php?id=1566327}.}, this flux ratio corresponds to an expected mass ratio $\sim 0.94$.

\begin{figure}[h!]
    \centering
    \includegraphics[width=1\linewidth]{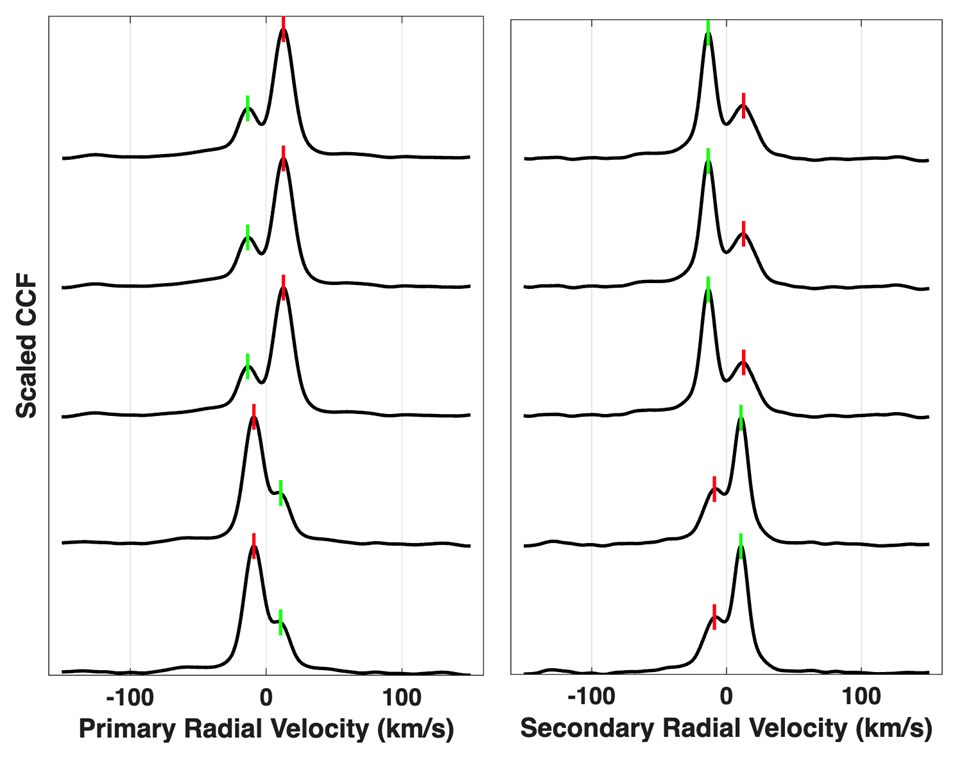}
    \caption{The 1D-cuts along the primary and secondary components radial velocities, of the 2D-CCF from JD-2458242.7673 (top) to JD-2458605.7720 (bottom).}
    \label{fig:todcor_3}
\end{figure}

Based on the optimized synthetic spectra determined above, we calculated the 2D-CCF of all the 5 spectra of HD 143811, removing beforehand the orders polluted by telluric lines. The  1D cuts of all the 2D-CCF along the primary and secondary velocities are shown in Figure~\ref{fig:todcor_3}. 

We derive relative radial velocities between the two epochs of $\Delta V_A=21.7$ km/s and $\Delta V_B=24.3$ km/s leading to $q=0.89$. This is in good agreement with the mass ratio inferred from the flux ratio above. Therefore, we allowed for $q \in [0.85, 0.95]$ in our analysis.

A solar metallicity, typical of young star-forming regions in the solar neighborhood \citep{d'orazi11,biazzo12}, is expected for Scorpius-Centaurus stars. However, in order to simultaneously account for the constraints on $E(B-V)$, age, and $q$ ratio, the metallicity of the star was allowed to vary between slightly subsolar and slightly super-solar values. 

All the constrains on the system detailed above were provided as input to \textsc{madys} \citep{squicciarini22}, so as to simultaneously derive the age of the system and the masses of the two components through isochrone fitting. As in \citetalias{squicciarini25}, we opted for PARSEC v2.0 \citep{nguyen22} isochrones. The analysis reveals a preference for $[\text{Fe/H}] = -0.13$ \citep[similarly to][]{zakhozhay_RVbinary_2022}, as the system age would be pushed to values older than 20 Myr or to smaller $q$ for $[\text{Fe/H}] = 0.00$. We derive an age $t=18 \pm 3$ Myr, masses $M_1=1.29 \pm 0.05 ~ M_\odot$ and $M_2=1.15 \pm 0.07 ~ M_\odot$, and temperatures $T_{\text{eff}, 1} = 6715 \pm 150$ K and $T_{\text{eff}, 2} = 6380 \pm 150$ K.

It is known that PARSEC isochrones tend to overestimate the effective temperature of young stars, but they typically provide a good match to the observed color–luminosity relations \citep[see, e.g,][]{ruizdern18,babusiaux18}; this gives us confidence that our mass determination is more robust than the effective temperature. In order to estimate the parameter uncertainty due to systematic differences across models, we repeated the analysis using MIST \citep{dotter16,choi16} and BHAC15 \citep{baraffe15} isochrones. MIST -- linearly interpolated between $\text{[Fe/H]} = -0.25$ and $\text{[Fe/H]} = 0.00$ -- yields similar results to PARSEC: $M_1=1.30 \pm 0.05 ~ M_\odot$, $M_2=1.17 \pm 0.07 ~ M_\odot$, $T_{\text{eff}, 1} = 6700 \pm 200$ K, $T_{\text{eff}, 2} = 6300 \pm 200$ K. Conversely, BHAC15 isochrones -- only available at solar metallicity -- are associated to higher masses and lower temperatures: $M_1=1.35 \pm 0.05 ~ M_\odot$, $M_2=1.21 \pm 0.07 ~ M_\odot$, $T_{\text{eff}, 1} = 6400 \pm 150$ K and $T_{\text{eff}, 2} = 6000 \pm 250$ K. However, the large $\chi^2$ of the best-fit solution ($\chi^2 > 10$), similar to the one found with PARSEC and MIST isochrones of solar metallicity, suggests that a slightly subsolar metallicity -- similar to the one found for the UCL member HD 139614 \citep{murphy21} -- might be a real feature of the system.

To summarize, we stress that the lack of a direct detection of the secondary, coupled with the intrinsic systematic uncertainties underlying models of pre-main sequence stars, does not allow us to determine in an unambiguous and accurate way the parameters of the system. Still, the total mass of the system ($M_{tot} \approx 2.50~M_\odot$) appears to be reasonably well constrained, and the temperature uncertainty does not impact in a significant way the derived planetary parameters (Appendix~\ref{app:physical_fit}).

\section{Ruling out the background scenario}
\label{app:proper_motion}

Confirming a DI candidate requires the assessment of its proper motion, that must be compatible with a bound orbit and significantly different from the motion of background stars that might appear by chance at a short projected separation from the target.

The relative proper motion of CC1 with respect to the star can be estimated as the difference between J2025 and J2016 astrometry, divided by their time span: $\Delta \mu_\alpha^* = 8.74 \pm 0.32$ mas/yr and $\Delta \mu_\delta = -2.53 \pm 0.34$ mas/yr for right ascension and declination, respectively. These correspond to an heliocentric proper motion $\mu_{\alpha, CC1}^* = -6.11 \pm 0.32$ mas/yr and $\mu_{\delta, CC1} = -27.51 \pm 0.34$ mas/yr. The total proper motion difference, $\Delta \mu = 9.10 \pm 0.32$ mas/yr, can be easily converted into a physical velocity difference: $\Delta v = 5.9 \pm 0.2$ km/s. A planet at the projected separation of CC1 cannot be at a physical separation from the central binary smaller than $d = \text{sep[mas]}/\varpi[mas] \approx 60$ au, $\varpi$ being the parallax of the system. If $\Delta v$ were larger than the escape velocity from the system at a distance $d$, then no closed orbit could be possible and CC1 would have to be a passing-by object. Being $v_{esc}(d) = \sqrt{2GM_{tot}/d} = 8.5 \pm 0.1$ km/s larger than $\Delta v$, the motion of CC1 is compatible with that of a bound object.

\begin{figure}[h!]
    \centering
    \includegraphics[width=1\linewidth]{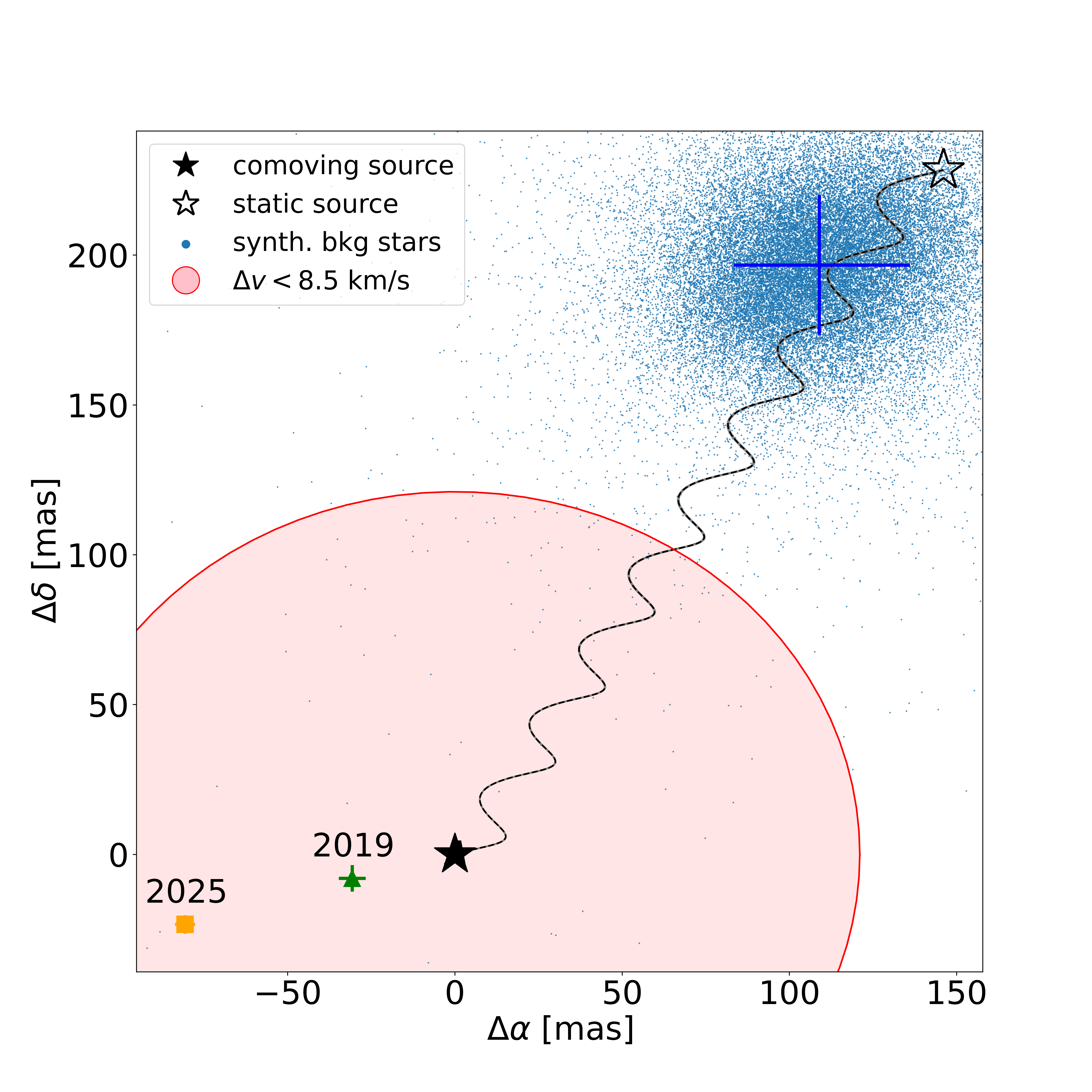}
    \caption{Astrometric displacements for CC1 in 2019 (green star) and 2025 (orange square) with respect to 2016. A static background star observed in 2025 would lie in the position of the empty star; a source with null relative motion compared to the target would lie where the filled star is. Simulated background sources from the Besançon model are shown as blue dots, with their median and ($16^{\text{th}}$, $84^{\text{th}}$) percentiles shown as a blue cross. The area encompassing bound orbits at $d=60$ au is filled in red.}
    \label{fig:proper_motion}
\end{figure}

In order to test the hypothesis that CC1 is a background interloper, we computed the probability that a background source be responsible for the observed astrometric shift \citep[see, e.g.,][]{squicciarini22_mu2sco}. We constructed therefore a sample of synthetic background stars using the Besançon Galaxy Model \citep{czekaj2014}. The sample is defined by a distance [0, 50] kpc, an apparent H magnitude similar to the one of CC1 ($H \in [19.5, 20.1]$ mag), and a radius = 1$^\circ$ centered on the position of HD 143811. The sample is composed of 38825 stars, whose median and [16, 84] percentiles of the proper motion components are $\mu_{\alpha, \text{gal}}^* = -4.02_{-2.77}^{+2.94}~\text{mas yr}^{-1}$ and $\mu_{\delta, \text{gal}} = -3.44_{-2.51}^{+2.53}~\text{mas yr}^{-1}$. These values are significantly different from those of CC1, which crucially exhibits a $\mu_\delta$ that is even more negative -- that is, even farther away from the bulk of the background star population -- than the one of HD 143811. Indeed, only 11 stars have a more extreme $\mu_\delta$ than the star and a total proper motion compatible with a bound orbit (Figure~\ref{fig:proper_motion}). This corresponds to a false alarm probability of $3 \times 10^{-4}$, that is, a C.L. = 3.6 $\sigma$. 

Additional information on the nature of the source is provided by its colors. We showed in \citetalias{squicciarini25} that the ($H2-H3$, $H2$) position of CC1 is not decisive, as it can be compatible with both a planet and a background star; however, the nondetection of the source in IFS data can be in principle a discriminating factor between the two scenarios. IFS detection limits correspond to an apparent $J>21.34$ mag; coupled with the measured H-band flux for CC1, this sets a lower limit to the $J-H2$ color of the source, $J-H2 > 1.4$ mag. While a red color like this is not atypical for young giant planets \citep[see, e.g., HR 8799 planets;][]{bonnefoy16}, we do not expect to measure it in background stars. Using the same synthetic sample as above, we find a median apparent $J=20.4 \pm 0.2$ mag, well within our detection capabilities; in fact, no source in the entire sample is dimmer than our detection limits -- the dimmest source having $J=20.9$ mag.

\section{Details on the orbital fitting}\label{app:orbitize}

\begin{figure}[h!]
    \centering
    \includegraphics[trim={0.5cm 0cm 0.8cm 0cm},clip,width=\linewidth]{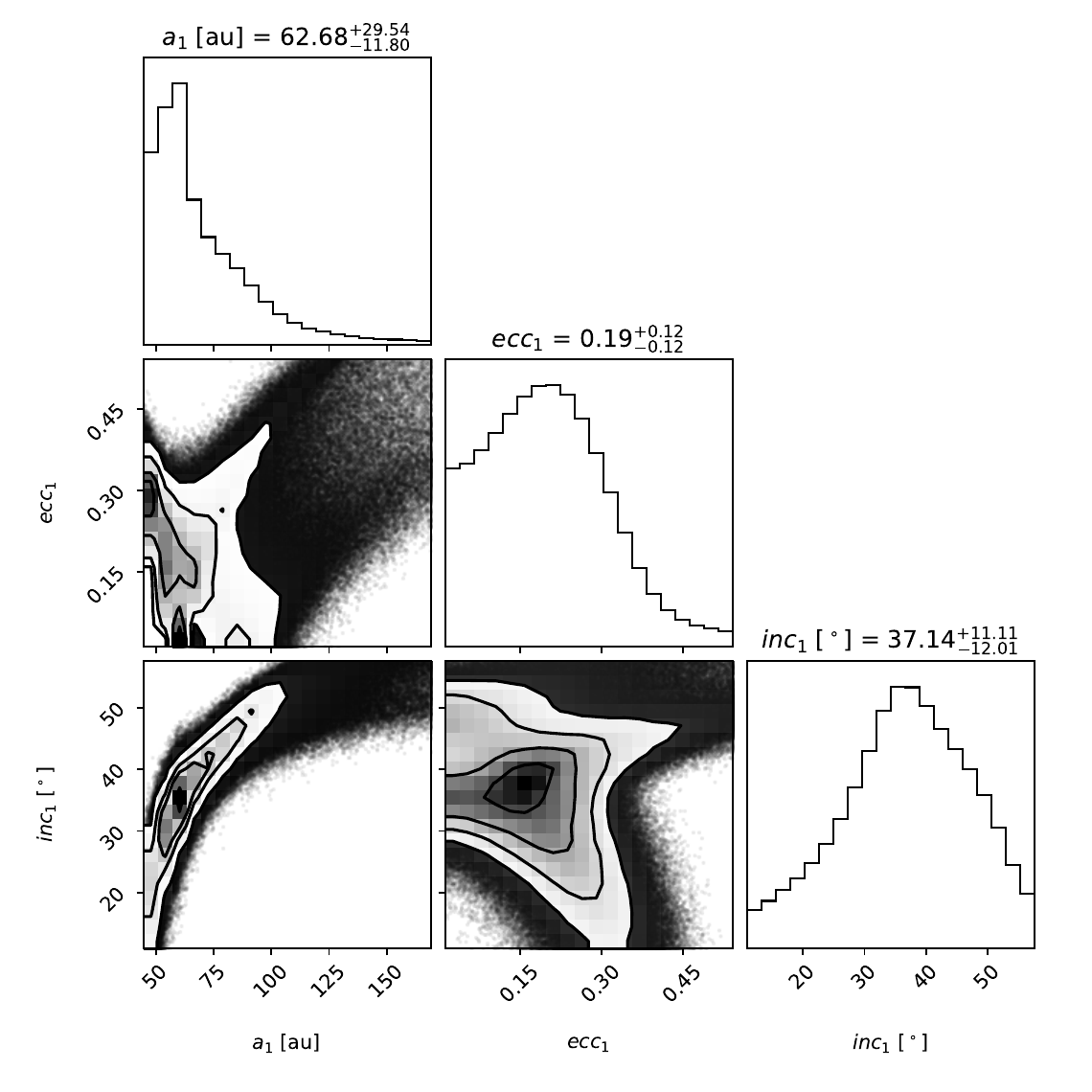}
    \caption{Posterior distributions of orbital parameters retrieved by OFTI ($10^{6}$ accepted orbits). We did not plot unconstrained parameters with mostly flat posterior distributions.}
    \label{fig:orbitizemcmc}
\end{figure}

The priors adopted for the orbital fitting of the planet astrometry are provided in Table~\ref{tab:inferred_params_orbit}. Default priors were used: a log-uniform prior for the semi-major axis; a sine-uniform distribution for the inclination; uniform priors for the other parameters (eccentricity, argument of periastron, position angle of nodes and epoch of periastron passage). The results of the analysis are presented in the same table, excluding those parameters that were not constrained by the analysis. The corner plot of the Bayesian analysis is shown in Figure~\ref{fig:orbitizemcmc}.

\begin{table}[h!]
\centering
\caption{Priors and inferred parameters from the MCMC analysis obtained with {\tt orbitize!}.}
\begin{tabular}{lcc}
\hline
\hline
 Orbital parameters & Prior & Final estimate \\
\hline
Semi-major axis (au) & [$10^{-3}$, $10^4$] (log) & $63 ^{+30}_{-12}$ \\
Eccentricity & [0,1]& $0.19 \pm 0.12$ \\ 
inclination ($^{\circ}$) & [0,90] (sine)  &$37 ^{+11}_{-12}$ \\
Period (yr) & & $319 ^{+248}_{-86}$\\
\hline
\end{tabular}
\tablefoot{The reported values are the median of the marginalized posterior distribution, with errors corresponding to the $16^{\text{th}}$ and $84^{\text{th}}$ percentiles.}
\label{tab:inferred_params_orbit}
\end{table}

\section{Derivation of planetary parameters}\label{app:physical_fit}

Since spectro-photometric measurements from DI instruments are expressed in units of source-to-star contrasts, the first step of the analysis of a planet spectrum deals with the conversion of contrast data into absolute fluxes. Compared to the standard approach (see \citetalias{squicciarini25}), where a synthetic stellar spectrum from the BT-Nextgen AGSS2009 library \citep{allard11} is used, we built an unresolved binary spectrum by summing up two such spectra with $\teff=6715$ K and $\teff=6380$ K (see Appendix~\ref{app:star_params}). This spectrum was then, as in the standard procedure, renormalized so as to match the measured H-band magnitude from 2MASS \citep{2mass}. We verified that the impact of the binary spectrum compared to the standard single-star spectrum on the planet flux is negligible ($\approx 0.2\%$) across GPI wavelengths, increasing to 4\% only in the case of the IFS upper limit. Even when accounting for the systematic uncertainties on the effective temperature, the impact on the upper limit remains limited ($< 7\%$). Hence, the spectral analysis is robust against uncertainties in the mass ratio $q=0.90 \pm 0.05$ and the effective temperatures of the two stars.

As mentioned in Section~\ref{sec:atmo_fit}, we analyzed six physical parameters using a Markov Chain Monte Carlo (MCMC) framework applied to a grid of coupled atmosphere–interior models from HADES \citep{wilkinson24}. We employ the affine-invariant MCMC ensemble sampler \texttt{emcee} to explore the parameter space. We use 50 walkers and run the chain for 400,000 steps after a conservative burn-in phase of 2,000 steps, ensuring the run is approximately 50 times the estimated autocorrelation time for the best-constrained parameters. We adopt uniform (flat) priors for all fitted parameters, given in Table \ref{tab:inferred_params_atmo}.

\begin{table}[h!]
\centering
\caption{Priors and inferred parameters from the MCMC analysis obtained with HADES codes.}
\begin{tabular}{lcc}
\hline
\hline
Planetary parameters & Prior & Final estimate \\
\hline
Mass (M$_p$/M$_{\text{Jup}}$) & $[0.1, 8.0]$ & $6.1^{+0.7}_{-0.9}$ \\
Intrinsic Temp. (T$_{\text{int}}$) [K] & $[300, 1500]$ & $1000 \pm 30$ \\
Metallicity ([Fe/H]) [dex] & $[-2, 3]$ & $0.1^{+0.8}_{-0.4}$ \\
Core Mass (M$_{\text{core}}$/M$_{\oplus}$) & $[1, 100]$ & Unconstrained\tablefootnote{The posterior for M$_{\text{core}}$ is bimodal and poorly constrained; see Fig. \ref{fig:HADES_corner}.} \\
Sedimentation (f$_{\text{sed}}$) & $[1, 6]$ & $2.8^{+1.1}_{-0.8}$ \\
Vertical Mixing ($\log K_{zz}$) & $[1, 12]$ & $9.0^{+1.2}_{-1.5}$ \\
Radius ($R_{\text{Jup}}$) & Model derived & $1.4 \pm 0.1$ \\
\hline
\end{tabular}
\tablefoot{All priors are uniform in the given ranges. The reported values are the median of the marginalized posterior distribution, with errors corresponding to the 16th and 84th percentiles.}
\label{tab:inferred_params_atmo}
\end{table}

The age of the system is used as an additional constraint.
For each step in the MCMC, the age is predicted from the physical parameters using the additional pre-computed HADES evolution model.
This predicted age is then compared to the known system upper age of $21$ Myr, which is treated as a soft upper limit. We use the upper bounds of the estimated median disk life for the stellar spectral type from \cite{pfalzner22} of $5$ Myr as a soft lower limit to the planets age. The age is estimated considering a medium entropy value per mass from \cite{mordasini17} equivalent to a "warm start".
The likelihood for the soft limits, as well as for the instrumental upper limit from the SPHERE-IFS data, is calculated using the cumulative distribution function of a Gaussian. This allows for uncertainty on the systems age estimation and for the noise in the SPHERE-IFS data.

\begin{figure*}
    \centering
    \includegraphics[width=\linewidth]{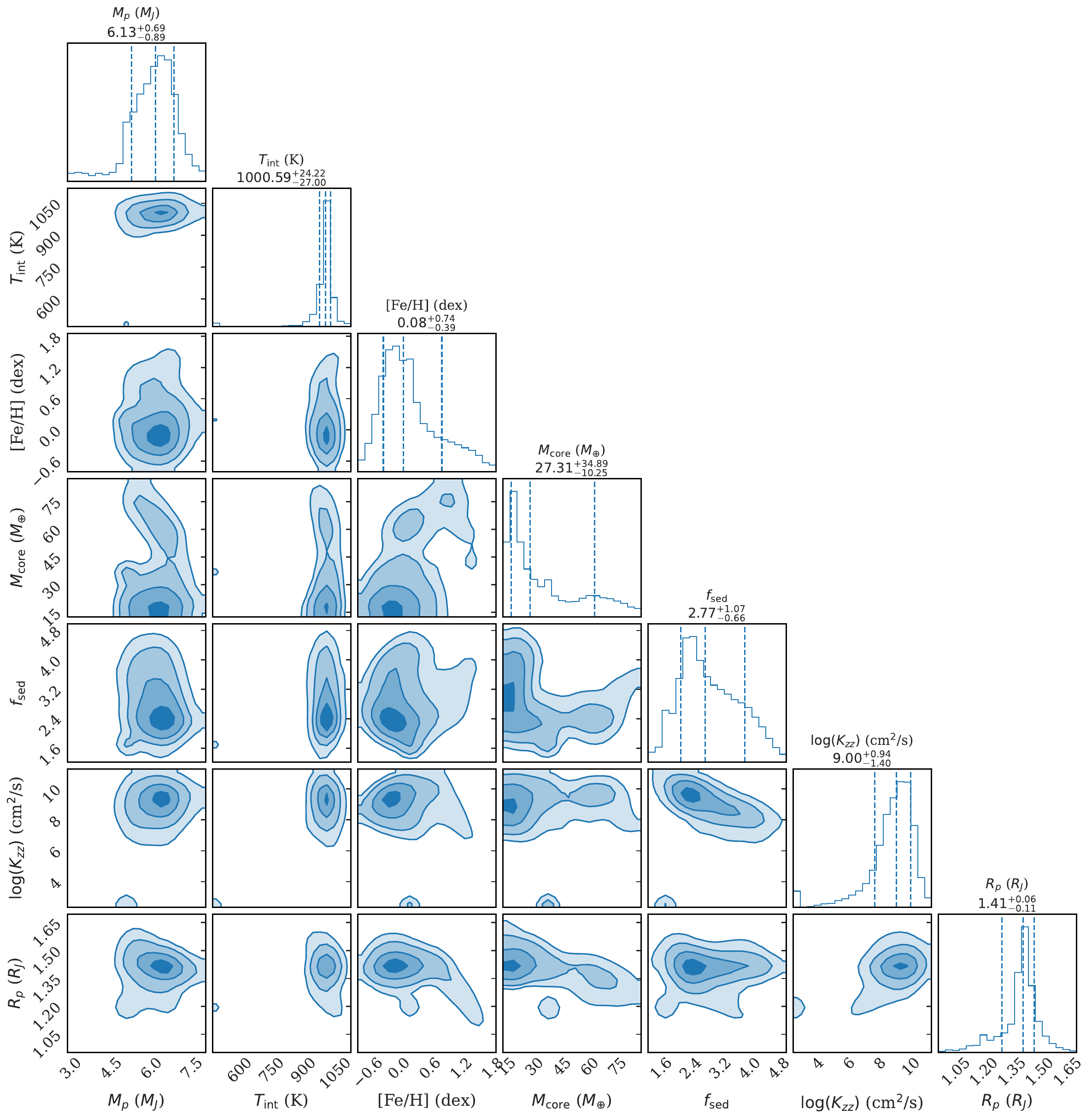}
    \caption{Posterior distributions of physical parameters retrieved using the HADES grid.}
    \label{fig:HADES_corner}
\end{figure*}

The derived posterior distributions are shown in Figure \ref{fig:HADES_corner}, with the inferred planetary parameters summarized in Table \ref{tab:inferred_params_atmo}.

\end{appendix}

\end{document}